\documentclass[12pt,preprint]{aastex}
\usepackage{epsfig}

\def\fd{_{\rm fd}}

\shorttitle{Frame-dragging and Galactic-Center stars}
\shortauthors{Rahul Kannan, Prasenjit Saha}
\begin{document}
\title{Frame-dragging and the kinematics of Galactic-Center stars}
\author{Rahul Kannan\altaffilmark{1,2}}
\and
\author{Prasenjit Saha\altaffilmark{2}}
\altaffiltext{1}{Department of Physics, Indian Institute of Technology,
Kharagpur 721 302, India}
\altaffiltext{2}{Institute for Theoretical Physics, University of Z\"{u}rich,
Winterthurerstrasse 190, CH-8057 Z\"{u}rich, Switzerland}
\begin{abstract}
  We calculate the effects of frame dragging on the Galactic-Center
  stars. Assuming the stars are only slightly relativistic, we derive
  an approximation to the Kerr metric, which turns out to be a weak
  field Schwarzschild metric plus a frame dragging term. By
  numerically integrating the resulting geodesic equations, we compute
  the effect on keplerian elements and the kinematics. We find that
  the kinematic effect at pericenter passage is proportional to
  $(a(1-e^2))^{-2}$. For known Galactic-center stars it is of order 10
  m/s. If observed this would provide a measurement of the spin of the
  black hole.
\end{abstract}


\section{Introduction}

The center of the Milky way is a very interesting region. It contains
a massive black hole (MBH) of $ \sim 3 \times 10^6 M_\odot$. The
central parsec contains thousands of stars . For a small but a growing
number of these, due to the relatively close proximity of the MBH and
short orbital periods, the orbital parameters have been accurately
measured
\citep{2003ApJ...596.1015S, 2005ApJ...620..744G, 2005ApJ...628..246E} .
Some of the stars have pericenter velocities as high as a few percent of $c$.
Hence, as shown in \cite{2006ApJ...639L..21Z} the general relativistic
effect of O($\beta^2$) should be observable.

At $O(\beta^3)$ general relativity predicts a new effect, which is
that a spinning black hole drags the surrounding space-time along with
it. Under the rotational frame-dragging effect (also known as
Lense-Thirring effect), the frame of reference with minimal time
dilation is one which is rotating around the object as viewed by a
distant observer. If this effect could be observed for GC stars, then
in principle the spin of the MBH can be measured. A method based on
the orbital dynamics of the GC stars is more direct than the usual
approach which requires modeling the effect of spin on the accretion
disk \citep[see for example][]{2008AIPC..968..265N}.

The effect of $O(\beta^3$) terms on the keplerian elements and on
astrometry were discussed by \cite{1998AcA....48..653J} and
\cite{2000ApJ...542..328F}. \cite{2008ApJ...674L..25W} goes on to
consider $O(\beta^4$) as well.  The resulting astrometric effects are
so small that they can only be observed on stars which are closer to
the MBH than the observed GC stars.

In this paper we concentrate on the effects of the $O(\beta^3)$ terms
on the kinematics of the GC stars.  Traditionally the effect of relativistic
perturbations on orbital dynamics have been studied either using
post-newtonian celestial mechanics \citep{1972gcpa.book.....W} or
pseudo-newtonian equations \citep{1999A&A...343..325S}. We adopt a
different and conceptually simpler approach. We do a low velocity
perturbative expansion of the Kerr metric and then numerically
integrate the resulting geodesic equations.

\section{The Model}

Our starting point is the Kerr metric in Boyer-Lindquist co-ordinates
\citep[see for example][]{1973grav.book.....M}:
\begin{equation}
ds^2 = -\frac{\Delta}{\rho^2}\left(dt -s\sin^2\theta\,d\phi\right)^2
+ \frac{\sin^2{\theta}}{\rho^2}\left((r^2 + s^2)d\phi - sdt\right)^2
+ \frac{\rho^2}{\Delta}dr^2 + {\rho^2}{d\theta^2}
\end{equation}
where
\begin{equation}
\Delta \equiv r^2-2Mr+s^2, \qquad
\rho^2 \equiv r^2+s^2\cos^2\theta
\end{equation} 
Here $s$ is the spin of the black hole and $M$ is the mass and  $G=c=1$.

The Kerr metric describes the space-time outside a rotating black
hole. The metric itself is complicated and is difficult to solve even
numerically. So we make some approximations to make the solution
easier.  We consider the case where
\begin{equation}
v^2 \sim 1/r
\label{approx}
\end{equation}
This is equivalent to saying that the system is not very relativistic.
To agree with Eq.~(\ref{approx}) we assume r is $O(\epsilon^{-2})$ and
$dr$, $rd\theta$ and $rd\phi$ are all $O(\epsilon)$.  We then replace
\begin{equation}
r \rightarrow \epsilon^{-2}r
\end{equation}
\begin{equation}
dr \rightarrow \epsilon dr
\end{equation}
\begin{equation}
d\theta \rightarrow \epsilon^3d\theta
\end{equation}
\begin{equation}
d\phi \rightarrow \epsilon^3d\phi
\end{equation}
in the Kerr metric, put $M=1$ and Taylor expand up to $O(\epsilon^5)$.
We get the following metric:
\begin{equation}
ds^2 = -\left(1-\frac{2\epsilon^2}{r}\right)dt^2
+ \left(1+\frac{2\epsilon^2}{r}\right)\epsilon^2dr^2
+ \epsilon^2r^2d\theta^2 + \epsilon^2r^2\sin^2{\theta}d\phi^2
- \frac{4s\epsilon^5}{r}\sin^2{\theta}dtd\phi 
\end{equation}
which is equivalent to a weak Schwarzschild field plus a
frame-dragging effect. The above metric is only valid for systems with
$\beta \ll 1$. In particular, it is not valid for null geodesics as
light does not satisfy Eq.~(\ref{approx}).

Applying the Euler-Lagrange equations, we get the following geodesic
equations for $t$, $r$, $\theta$ and $\phi$.
\begin{equation}
\left(1-\frac{2\epsilon^2}{r}\right)\ddot{t}
+\frac{2\epsilon^2}{r^2}\dot{r}\dot{t}+O(\epsilon^5) = 0
\label{eq-geo1}
\end{equation}
\begin{equation}
\left(1+\frac{2\epsilon^2}{r}\right)\ddot{r}=
r\dot{\theta}^2+r\sin^2{\theta}\dot{\phi}^2-\frac{\dot{t}^2}{r^2}
+\epsilon^2\frac{\dot{r}^2}{r^2}
+\frac{2s\epsilon^3}{r^2}\sin^2{\theta}\dot{t}\dot{\phi}
\end{equation}
\begin{equation}
r^2\ddot{\theta}+2r\dot{r}\dot{\theta}=
-\frac{2a\epsilon^3}{r}\sin{2\theta}\dot{t}\dot{\phi}
+\frac{r^2\sin{2\theta}\dot{\phi}^2}{2}
\end{equation}
\begin{equation}
r^2\sin^2{\theta}\ddot{\phi}+r^2\sin{2\theta}\dot{\theta}\dot{\phi}
+2r\sin^2{\theta}\dot{r}\dot{\phi}+\frac{2s\epsilon^3}{r}
\left(\frac{\sin^2{\theta}}{r}\dot{r}\dot{t}
-\sin{2\theta} \dot{\theta}\dot{t}-\sin^2{\theta}\ddot{t}\right)=0
\label{eq-geo4}
\end{equation}
The leading order terms give the Newtonian equations, terms up to
$O(\epsilon^2)$ gives Schwarzschild equations and terms up to
$O(\epsilon^3)$ gives the frame-dragging effect. We remark that
$\epsilon$ as used here is just a label for keeping track of orders.
For numerical work, we set $\epsilon=1$.

\section{Results}

We now proceed with the numerical integration of the geodesic
equations (\ref{eq-geo1}--\ref{eq-geo4}). For simplicity we choose the
unit of length to be the gravitational radius $GM/c^2$ of the
central MBH ($\sim 5 \times 10^6$ km) and velocity to be in units of
speed of light, which makes the unit of time to be around $17s$.

We choose $a$, $e$, $I$, $\Omega$ and $\omega$ as the free parameters,
where $a$ is the semi-major axis of the ellipse, $e$ the eccentricity,
$I$ the inclination with respect to the spin axis of the black hole,
$\Omega$ the longitude of ascending node and $\omega$ is the argument
of perihelion. We start the integration at apocenter and integrate
from $t=0$ to $t=2\pi a^{3/2}$ which is exactly equal to one newtonian
orbital period. We then re-compute $a,e,I,\Omega,\omega$ at the
end of one integration and hence find the effect on the keplerian
elements. Relevant formulas for the initial conditions are given in
the Appendix.

We actually integrate three different sets of geodesic equations:
\begin{itemize}
\item The equations for the newtonian case (includes only zero order
  $\epsilon$ terms). The orbital elements are completely unchanged.
\item The Schwarzschild case (terms up to $\epsilon^2$). There is a
  pericenter shift in this case which we call $\Delta\omega_s$.
\item The frame dragging case (terms up to $\epsilon^3$). There is a
  shift in the node $\Delta\Omega\fd$ and a further shift
  $\Delta\omega\fd$. There is no net effect in $a$, $e$ and $I$.
\end{itemize}
We have also computed the velocity differences between these cases.

We examine the numerical data to verify the known parameter
dependencies of the various relativistic effects or find an empirical
formula. All the effects turn out to depend upon
\begin{equation}
p = a(1-{e^2})
\label{pdef}
\end{equation}
which is the square of the angular momentum in gravitational units.

\begin{figure}
\epsscale{.80}
\plotone{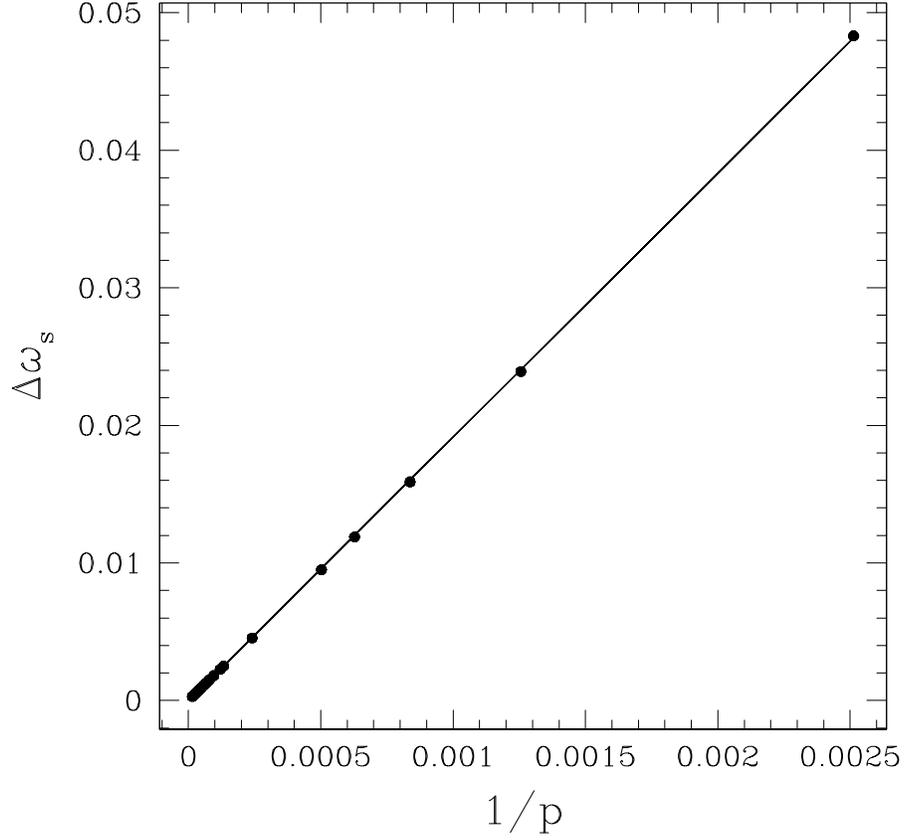}
\caption{Pericenter precession $\Delta\omega_s$ due to $O(\epsilon^2)$
terms, plotted against $1/p$, for a sample of orbits with different
$a,e,I$. The line shows the classical result of Eq.~(\ref{ws}). }
\label{sch}
\end{figure}

First the well known expression for the pericenter shift
\begin{equation}
\Delta \omega_s = \frac{6\pi}{p}   
\label{ws}
\end{equation}
 has been verified numerically as shown in Fig.~\ref{sch}.

\begin{figure}
\epsscale{.55}
\plotone{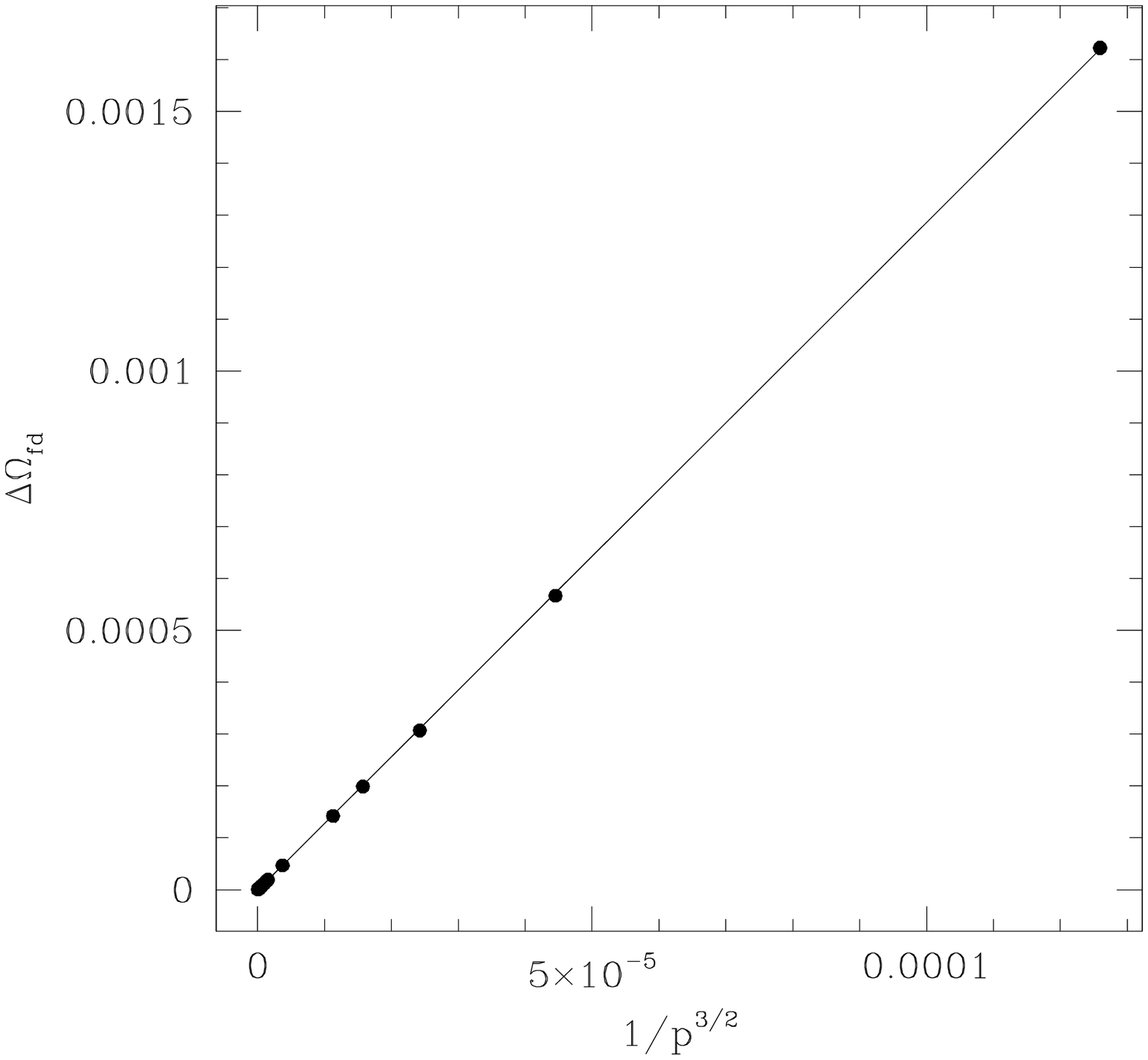}
\plotone{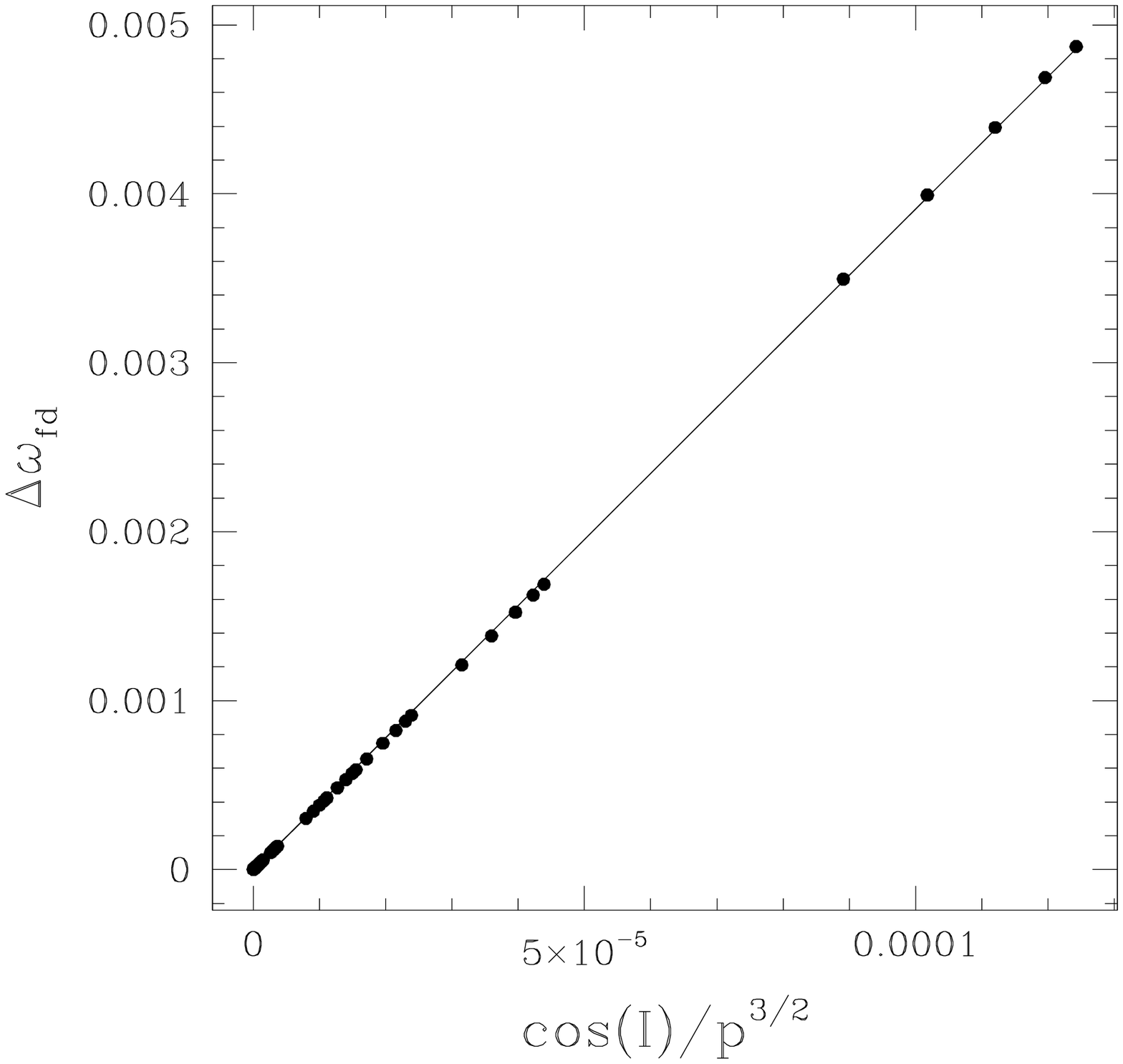}
\caption{Frame-dragging precession $\Delta\Omega\fd$ and
$\Delta\omega\fd$ due to $O(\epsilon^3)$ terms,plotted against the
dependencies (\ref{Wfd}) and (\ref{wfd}) respectively, for a range of
orbits as in Fig.~\ref{sch}}
\label{fdangle}
\end{figure}

Next we consider the frame-dragging precession, for which we
empirically infer:
\begin{equation}
\Delta \Omega\fd = \frac{4\pi s}{p^{3/2}}
\label{Wfd}
\end{equation}
and
\begin{equation}
\Delta \omega\fd = -12\pi s\cos I/p^{3/2}
\label{wfd}
\end{equation}
as Fig \ref{fdangle} verifies. For a plane perpendicular to the spin
axis of the black hole ($\cos I=1$), $\Delta\Omega\fd+\Delta\omega\fd$
corresponds to the total pericenter precession which we see is equal
to $-8\pi p^{-3/2}$ \citep[cf. Section 3 of][]{2005ApJ...622..878W}.
The physical shift of the apocentre is $\simeq a(1+e)\Delta\Omega\fd$.
As viewed from a distance of 8 kpc this translates into an astrometric
shift of
\begin{equation}
\Delta\alpha = a(1+e)\Delta\Omega\fd \times 4\mu {\rm as}
\label{alph}
\end{equation}

It should be noted that we have calculated the change in keplerian
elements after one complete revolution around the MBH.  Dividing by
the orbital period $2\pi a^{3/2}$ gives the mean rate of change.
Doing so in Eqs.~\ref{ws}, \ref{Wfd} and \ref{wfd} gives expressions
matching Eqs.~6, 4 and 5 of \cite{1998AcA....48..653J}.

\begin{figure}
\epsscale{.45}
\plotone{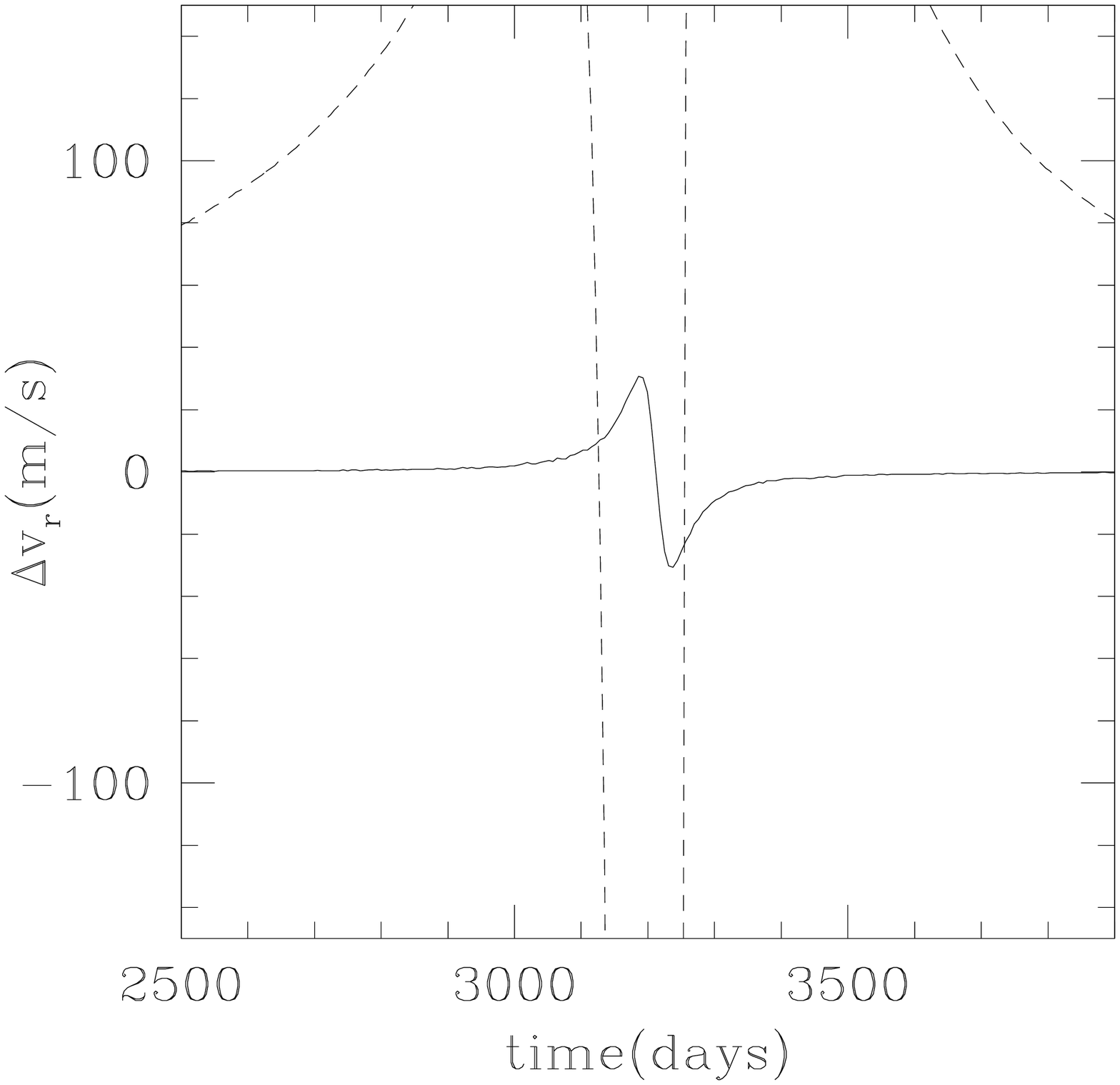}
\plotone{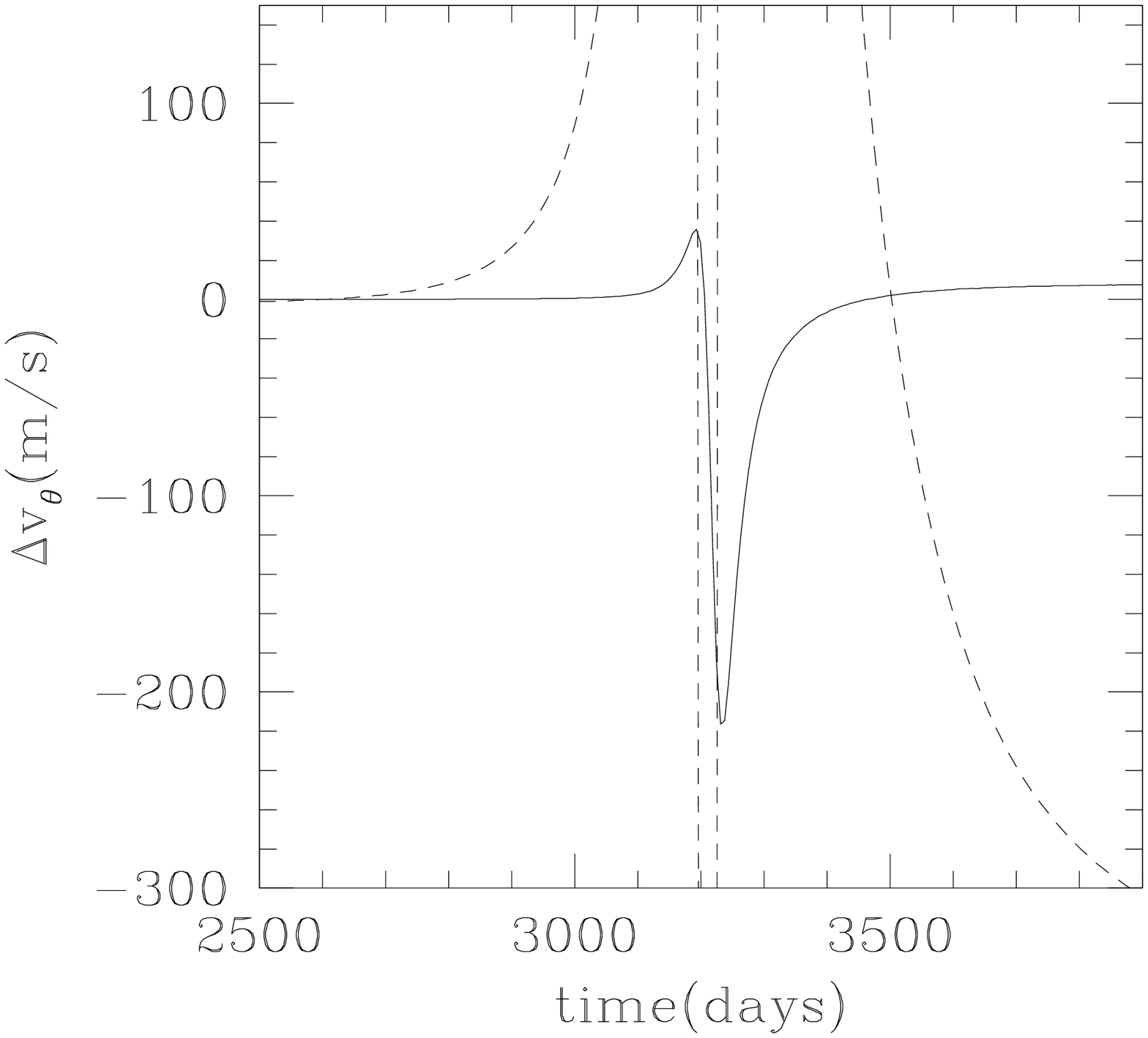}
\plotone{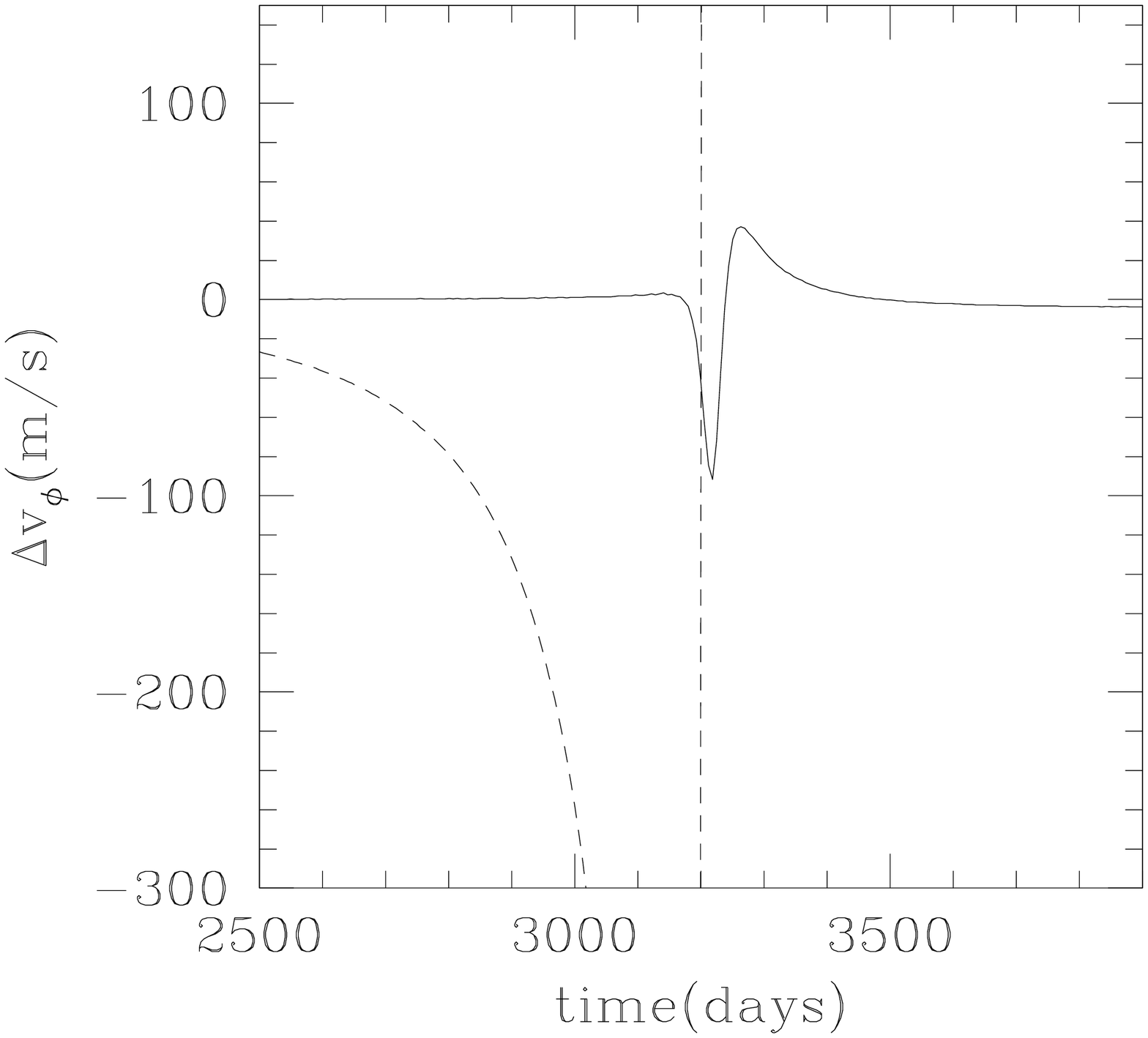}
\caption{Comparison between the Schwarzschild (dashed curves) and
frame-dragging (solid curves) contributions to the three velocity
components $v_r$, $v_\theta$ and $v_\phi$. The time starts from the
apocenter. For this orbit $a=3\times 10^4$, $e=0.90$ and $I=0.5$.}
\label{kincomp}
\end{figure}

\begin{figure}
\epsscale{.55}
\plotone{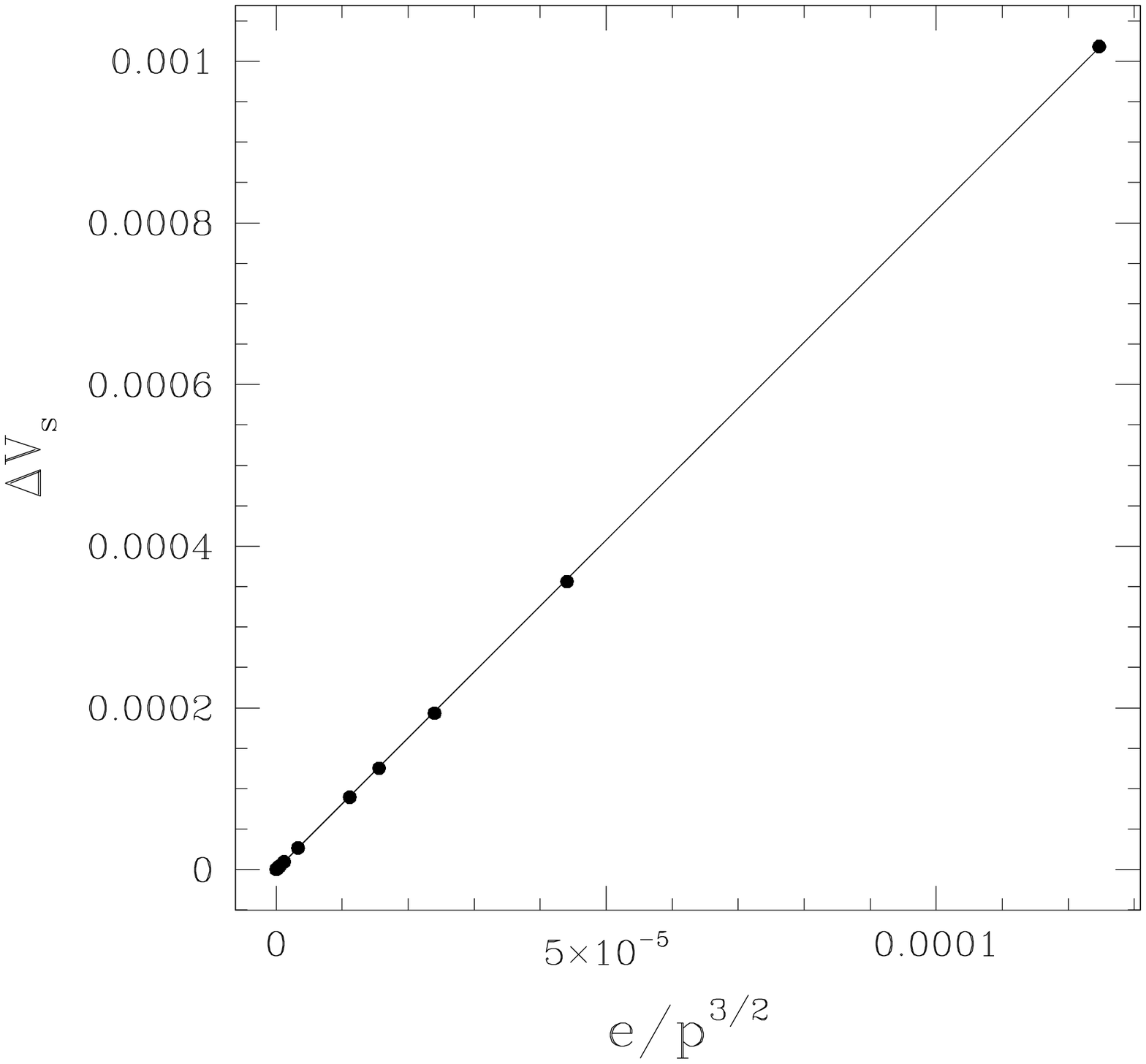}
\plotone{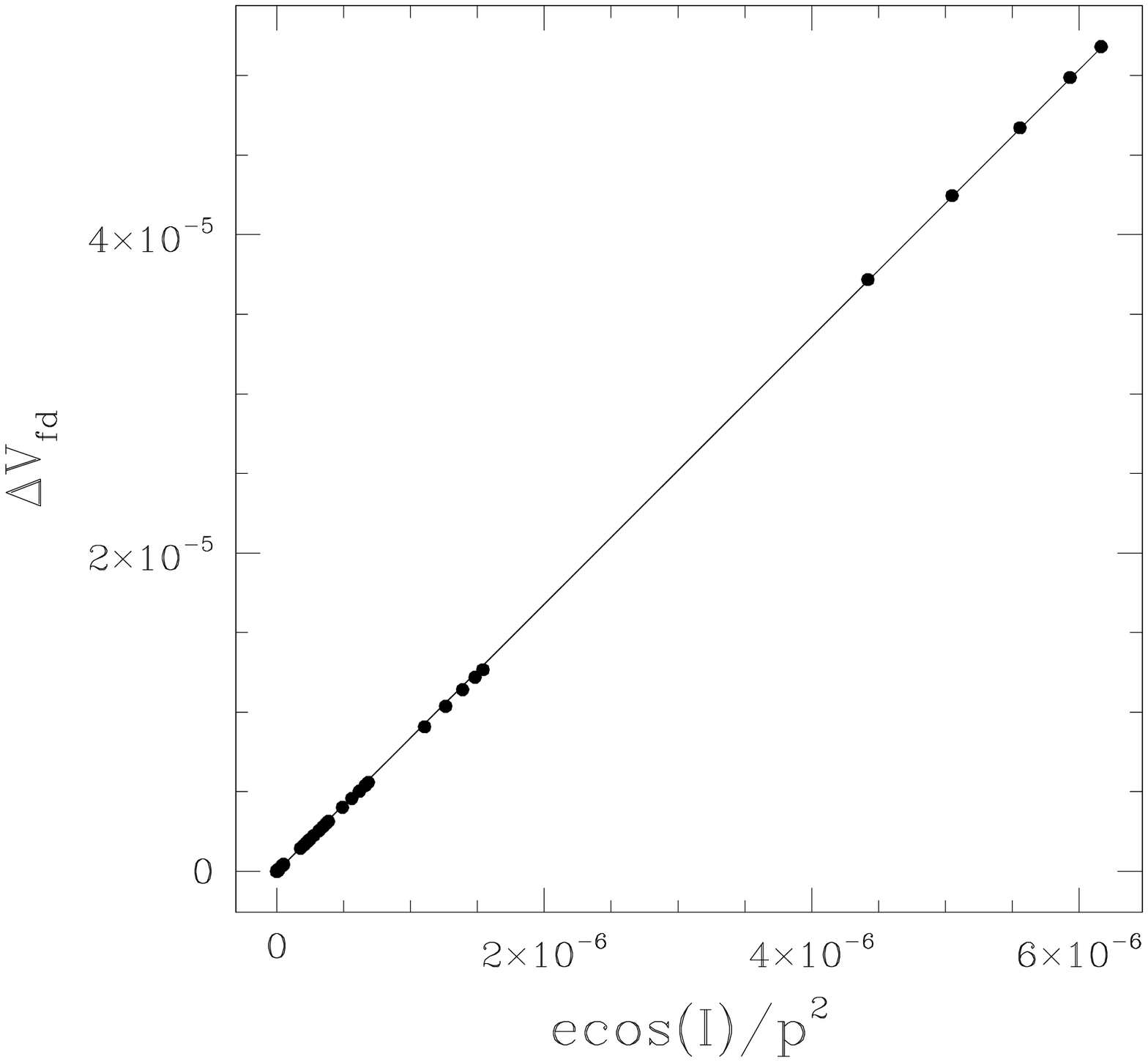}
\caption{Peak kinematic contributions $\Delta V_s$ and $\Delta V\fd$,
  plotted against the dependencies (\ref{vs}) and (\ref{vfd})
  respectively.}
\label{kin}
\end{figure}

We now consider kinematic effects. Fig.~\ref{kincomp} shows the
comparison between the Schwarzschild and frame-dragging effects on the
three velocity components. The effect is a fraction of a kilometer per
second during pericenter passage for typical GC star orbit parameters.
Note that the velocity components $v_r = \dot{r}$, $v_\theta =
r\dot{\theta}$, $v_\phi = r\sin\theta \dot{\phi}$ are actually
derivatives with respect to the conformal parameter.  Finally, we
consider the maximum velocity difference between the frame-dragging
and Schwarzschild effects ($\Delta V\fd$) and between Schwarzchild and
Newtonian effects ($\Delta V_{s}$). Fig.~\ref{kin} shows the empirical
relations:
\begin{equation} 
\Delta V_s \approx \frac{8e}{p^{3/2}}
\label{vs}
\end{equation}
\begin{equation}
\Delta V\fd \approx \frac{-8.4es\cos I}{p^2}
\label{vfd}
\end{equation} 

All these empirical relations are fairly accurate for parameters
typical of GC stars but not exact. On the figures the points do not
exactly lie on the line which suggests that these are leading order
effects.

Table 1 gives the values of the various relativistic effects according
to our empirical formulas for a sample of GC stars and two binary pulsars.
For the GC stars we estimated $p$ using Eq.~(\ref{pdef}),
from the $a$ and $e$ values tabulated in \cite{2005ApJ...628..246E}.
For the binary pulsars we derived $p$ using Eq.~(\ref{ws}) and the
precession rate given in \cite{lrr-2006-3}. We see from the table that
the GC stars are more relativistic than binary pulsars. The advantage
is that the binary pulsars have very short orbital periods (less than
a day) and hence we get many more orbits.

\section{Conclusions}

We see that the maximum kinematic effect in known GC stars $\Delta
V\fd$ is of the order of a few 10's of m/s during the few weeks around
the pericenter passage. Although this level of accuracy is difficult
to achieve for GC stars, it is not implausible. Extrasolar planet
searches regularly reach an accuracy better than 1 m/s
\citep{2006Natur.441..305L} and new technologies for radial velocity
measurements may be able to obtain precision as high as 1 cm/s
\citep{2008Natur.452..610L}.

There are also two theoretical problems which remain to be solved.

First, an accurate calculation of the redshift as a function of time
is required (as it is the observable quantity), rather than velocity
as a function of time as calculated here.  Both the kinematic and
gravitational redshifts are involved.  We do not know of any
approximate method for calculating the redshift in this case, as our
approximate metric is not valid for light.  It may be necessary to
calculate null geodesics in the full Kerr metric.

Second, the relativistic effects have to be separated from the
newtonian effects of other masses, such as nearby stars, gas and
dark-matter clouds.  These could overshadow the frame-dragging
contribution, especially since some newtonian dynamical processes in
the GC region can be unexpectedly strong because of resonances
\citep{2007MNRAS.379.1083G,2008ApJ...683L.151L}.  However, newtonian
perturbations from other masses would not give the distinctive time
dependence in the kinematics that frame-dragging does
(Fig.~\ref{kincomp}). Hence, we can be optimistic about dis-entangling
frame-dragging from all the newtonian effects.

\newpage

\appendix \section{Evaluating the orbital elements}

The numerical integrations in this paper are done in Boyer-Lindquist
coordinates, whereas the results are presented in terms of Keplerian
orbital elements.  To convert between them, we use standard relations
from celestial mechanics.  In practice, we only need to use the
classical formulas at or near the apocenter, so we will treat
$r,\theta,\phi$ as ordinary spherical polar coordinates.

For the initial conditions, we need to set up a star at apocenter with
given $a,e,I,\Omega,\omega$. Since the relativistic effects are
minimal here we set $\dot{t} = 1$. We start by defining a temporary
cartesian coordinate system, centered at the black hole, but oriented
such that the star is on the $x$ axis with velocity along $+y$.  In
other words, we put the star at
\begin{equation}
   \pmatrix{r_{\rm apo} \cr 0 \cr 0}  \qquad
   \pmatrix{0 \cr v_{\rm apo} \cr 0}
\end{equation}
where
\begin{equation}
   r_{\rm apo} = a(1+e) \qquad
   v_{\rm apo} = \sqrt{1-e\over1+e} \, \sqrt{1/a}
\end{equation}
Applying the rotation
\begin{equation}
   R_z(\Omega) \, R_x(I) \, R_z(\omega+\pi)
\end{equation}
(right operator first) gives the position and velocity in the
reference cartesian system.  We then convert to spherical polar
coordinates.

For the inverse process at the end of an integration, we start by
computing position $\bf r$ and velocity $\bf v$ in cartesian
coordinates.  We then compute the specific angular momentum $\bf h$
and the Runge-Lenz vector $\bf e$.
\begin{equation}
   {\bf h} = {\bf r} \times {\bf v}  \qquad
   {\bf e} = {\bf v} \times {\bf h} - {{\bf r}\over r}
\end{equation}
The inclination $I$ and the longitude of the ascending node $\Omega$
are simply a way of specifying the orbital plane, and we have
\begin{equation}
\Omega = \arctan(h_y,h_x) + {\pi\over 2} \qquad
     I = \arctan\left(\sqrt{h_x^2+h_y^2},h_z\right)
\end{equation}
Here $\arctan$ means the two argument form also called {\tt atan2}.
For the argument of the perihelion, we consider the Runge-Lenz vector
(which is a vector having magnitude $e$ and pointing towards the
pericenter) in the orbital plane
\begin{equation}
\pmatrix{e\cos\omega \cr e\sin\omega \cr 0} =
         R_x(-I) \, R_z(-\Omega) \, {\bf e}
\end{equation}
and the left hand side gives $\omega$ and $e$.

\newpage

\bibliographystyle{apj}
\bibliography{ms.bbl}

\begin{thebibliography}{17}
\expandafter\ifx\csname natexlab\endcsname\relax\def\natexlab#1{#1}\fi

\bibitem[{{Eisenhauer} {et~al.}(2005){Eisenhauer}, {Genzel}, {Alexander},
  {Abuter}, {Paumard}, {Ott}, {Gilbert}, {Gillessen}, {Horrobin}, {Trippe},
  {Bonnet}, {Dumas}, {Hubin}, {Kaufer}, {Kissler-Patig}, {Monnet},
  {Str{\"o}bele}, {Szeifert}, {Eckart}, {Sch{\"o}del}, \&
  {Zucker}}]{2005ApJ...628..246E}
{Eisenhauer}, F., {Genzel}, R., {Alexander}, T., {Abuter}, R., {Paumard}, T.,
  {Ott}, T., {Gilbert}, A., {Gillessen}, S., {Horrobin}, M., {Trippe}, S.,
  {Bonnet}, H., {Dumas}, C., {Hubin}, N., {Kaufer}, A., {Kissler-Patig}, M.,
  {Monnet}, G., {Str{\"o}bele}, S., {Szeifert}, T., {Eckart}, A.,
  {Sch{\"o}del}, R., \& {Zucker}, S. 2005, \apj, 628, 246

\bibitem[{{Fragile} \& {Mathews}(2000)}]{2000ApJ...542..328F}
{Fragile}, P.~C. \& {Mathews}, G.~J. 2000, \apj, 542, 328

\bibitem[{{Ghez} {et~al.}(2005){Ghez}, {Salim}, {Hornstein}, {Tanner}, {Lu},
  {Morris}, {Becklin}, \& {Duch{\^e}ne}}]{2005ApJ...620..744G}
{Ghez}, A.~M., {Salim}, S., {Hornstein}, S.~D., {Tanner}, A., {Lu}, J.~R.,
  {Morris}, M., {Becklin}, E.~E., \& {Duch{\^e}ne}, G. 2005, \apj, 620, 744

\bibitem[{{G{\"u}rkan} \& {Hopman}(2007)}]{2007MNRAS.379.1083G}
{G{\"u}rkan}, M.~A. \& {Hopman}, C. 2007, \mnras, 379, 1083

\bibitem[{{Jaroszynski}(1998)}]{1998AcA....48..653J}
{Jaroszynski}, M. 1998, Acta Astronomica, 48, 653

\bibitem[{{Li} {et~al.}(2008){Li}, {Benedick}, {Fendel}, {Glenday},
  {K{\"a}rtner}, {Phillips}, {Sasselov}, {Szentgyorgyi}, \&
  {Walsworth}}]{2008Natur.452..610L}
{Li}, C.-H., {Benedick}, A.~J., {Fendel}, P., {Glenday}, A.~G., {K{\"a}rtner},
  F.~X., {Phillips}, D.~F., {Sasselov}, D., {Szentgyorgyi}, A., \& {Walsworth},
  R.~L. 2008, \nat, 452, 610

\bibitem[{{L{\"o}ckmann} {et~al.}(2008){L{\"o}ckmann}, {Baumgardt}, \&
  {Kroupa}}]{2008ApJ...683L.151L}
{L{\"o}ckmann}, U., {Baumgardt}, H., \& {Kroupa}, P. 2008, \apjl, 683, L151

\bibitem[{{Lovis} {et~al.}(2006){Lovis}, {Mayor}, {Pepe}, {Alibert}, {Benz},
  {Bouchy}, {Correia}, {Laskar}, {Mordasini}, {Queloz}, {Santos}, {Udry},
  {Bertaux}, \& {Sivan}}]{2006Natur.441..305L}
{Lovis}, C., {Mayor}, M., {Pepe}, F., {Alibert}, Y., {Benz}, W., {Bouchy}, F.,
  {Correia}, A.~C.~M., {Laskar}, J., {Mordasini}, C., {Queloz}, D., {Santos},
  N.~C., {Udry}, S., {Bertaux}, J.-L., \& {Sivan}, J.-P. 2006, \nat, 441, 305

\bibitem[{{Misner} {et~al.}(1973){Misner}, {Thorne}, \&
  {Wheeler}}]{1973grav.book.....M}
{Misner}, C.~W., {Thorne}, K.~S., \& {Wheeler}, J.~A. 1973, {Gravitation} (San
  Francisco: W.H.~Freeman and Co., 1973)

\bibitem[{{Narayan} {et~al.}(2008){Narayan}, {McClintock}, \&
  {Shafee}}]{2008AIPC..968..265N}
{Narayan}, R., {McClintock}, J.~E., \& {Shafee}, R. 2008, in American Institute
  of Physics Conference Series, Vol. 968, Astrophysics of Compact Objects, ed.
  Y.-F. {Yuan}, X.-D. {Li}, \& D.~{Lai}, 265--272

\bibitem[{{Sch{\"o}del} {et~al.}(2003){Sch{\"o}del}, {Ott}, {Genzel}, {Eckart},
  {Mouawad}, \& {Alexander}}]{2003ApJ...596.1015S}
{Sch{\"o}del}, R., {Ott}, T., {Genzel}, R., {Eckart}, A., {Mouawad}, N., \&
  {Alexander}, T. 2003, \apj, 596, 1015

\bibitem[{{Semer{\'a}k} \& {Karas}(1999)}]{1999A&A...343..325S}
{Semer{\'a}k}, O. \& {Karas}, V. 1999, \aap, 343, 325

\bibitem[{{Weinberg} {et~al.}(2005){Weinberg}, {Milosavljevi{\'c}}, \&
  {Ghez}}]{2005ApJ...622..878W}
{Weinberg}, N.~N., {Milosavljevi{\'c}}, M., \& {Ghez}, A.~M. 2005, \apj, 622,
  878

\bibitem[{{Weinberg}(1972)}]{1972gcpa.book.....W}
{Weinberg}, S. 1972, {Gravitation and Cosmology: Principles and Applications of
  the General Theory of Relativity} (Gravitation and Cosmology: Principles and
  Applications of the General Theory of Relativity, by Steven Weinberg,
  pp.~688.~ISBN 0-471-92567-5.~Wiley-VCH , July 1972.)

\bibitem[{Will(2006)}]{lrr-2006-3}
Will, C.~M. 2006, Living Reviews in Relativity, 9

\bibitem[{{Will}(2008)}]{2008ApJ...674L..25W}
{Will}, C.~M. 2008, \apjl, 674, L25

\bibitem[{{Zucker} {et~al.}(2006){Zucker}, {Alexander}, {Gillessen},
  {Eisenhauer}, \& {Genzel}}]{2006ApJ...639L..21Z}
{Zucker}, S., {Alexander}, T., {Gillessen}, S., {Eisenhauer}, F., \& {Genzel},
  R. 2006, \apjl, 639, L21

\end{thebibliography}

\newpage

\newpage

\newpage

\newpage

\newpage
 
\begin{deluxetable}{ccrrrrrrrrrrrcrl}
\tabletypesize{\tiny}
\tablecaption{Relativistic effects in GC stars and Binary Pulsars}
\tablewidth{0pt}
\tablehead{
\colhead{Object} & \colhead{p}
& \colhead{${\frac{\Delta \Omega\fd}{s} \atop \rm (radians)}$}
& \colhead{${\frac{\Delta \omega\fd}{s\cos I} \atop \rm (radians)}$}
& \colhead{${\frac{\Delta V\fd}{s\cos I} \atop \rm (km/s)}$}
& \colhead{${\Delta \omega_s \atop \rm (radians)}$}
& \colhead{${\Delta V_s \atop \rm (km/s)}$}
& \colhead{${\rm Astrometric\ shift \atop (\mu arcsec)}$}
}
\startdata
S1 & 8.6$\times 10^4$ & 5.0$\times 10^{-7}$ & -1.5$\times 10^{-6}$ & -1.2$\times 10^{-4}$ & 2.2$\times 10^{-4}$ & 0.034 & 0.27 \\
S2 & 6.8$\times 10^3$ & 2.2$\times 10^{-5}$ & -6.7$\times 10^{-5}$ & -4.7$\times 10^{-2}$ & 2.8$\times 10^{-3}$ & 3.7 & 4.9 \\
S8 & 1.1$\times 10^4$ & 1.1$\times 10^{-5}$ & -3.2$\times 10^{-5}$ & -1.9$\times 10^{-2}$ & 1.7$\times 10^{-3}$ & 1.9 & 6.5 \\
S12 & 1.3$\times 10^4$ & 8.7$\times 10^{-6}$ & -2.6$\times 10^{-5}$ & -1.4$\times 10^{-2}$ & 1.5$\times 10^{-3}$ & 1.5 & 4.5 \\
S13 & 4.4$\times 10^4$ & 1.4$\times 10^{-6}$ & -4.1$\times 10^{-6}$ & -5.1$\times 10^{-4}$ & 4.3$\times 10^{-4}$ & 0.10 & 0.40 \\
S14 & 6.4$\times 10^3$ & 2.5$\times 10^{-5}$ & -7.4$\times 10^{-5}$ & -5.8$\times 10^{-2}$ & 3.0$\times 10^{-3}$ & 4.4 & 10. \\
B1913+16 & 2.9$\times 10^5$ & 8.1$\times 10^{-8}$ & -2.4$\times 10^{-7}$ & -1.9$\times 10^{-5}$ & 6.5$\times 10^{-5}$ & 9.5$\times 10^{-3}$ & - \\
PSR J07370 & 2.3$\times 10^5$ & 1.2$\times 10^{-7}$ & -3.5$\times 10^{-7}$ & -4.3$\times 10^{-6}$ & 8.3$\times 10^{-5}$ & 1.9$\times 10^{-3}$ & - \\
\enddata
\end{deluxetable}

\newpage

\end{document}